 \documentclass[twocolumn,twocolappendix]{aastex631}

\usepackage{multirow}
\usepackage{enumitem}
\usepackage{newtxtext,newtxmath,ulem}
\usepackage[T1]{fontenc}
\usepackage{amsmath}
\usepackage{xcolor}

\def\be{\begin{equation}}
\def\ee{\end{equation}}
\def\ba{\begin{eqnarray}}
\def\ea{\end{eqnarray}}

\def\ltsima{$\; \buildrel < \over \sim \;$}
\def\simlt{\lower.5ex\hbox{\ltsima}}
\def\gtsima{$\; \buildrel > \over \sim \;$}
\def\simgt{\lower.5ex\hbox{\gtsima}}

\definecolor{falured}{rgb}{0.5, 0.09, 0.09}

\shorttitle{Super-virial gas in emission}
\shortauthors{Bisht et al.}

\begin{document}

\title{On the origin of the  $10^7$ K hot emitting gas in the Circumgalactic medium of the Milky Way}

\correspondingauthor{Mukesh Singh Bisht}
\email{msbisht@rrimail.rri.res.in}

\author[0000-0002-1497-4645]{Mukesh Singh Bisht}
\affiliation{Raman Research Institute,
Bengaluru - 560080, INDIA}

\author[0000-0003-1922-9406]{Biman B. Nath}
\affiliation{Raman Research Institute,
Bengaluru - 560080, INDIA}

\author[0000-0002-4822-3559]{Smita Mathur}
\affiliation{Astronomy Department, Ohio State University, 140 Wet 18th Avenue, Columbus, OH 43210, USA}
\affiliation{Center for Cosmology and Astro-particle Physics, Ohio State University, 191 West Woodruff Avenue, Columbus, OH 43210, USA}



\begin{abstract}
The presence of the $\approx 10^6$ K gas in the circumgalactic medium of the Milky Way has been well established. However, the location and the origin of the newly discovered hot gas at `super-virial' temperatures of $\approx 10^7$ K have been puzzling. This hot gas has been detected in both absorption and emission; here we focus on the emitting gas only. We show that both the `virial' and the `super-virial' temperature gas as observed in \emph{emission} occupy disk-like extraplanar regions, in addition to the diffuse virial temperature gas filling the halo of the Milky Way. We perform idealized hydrodynamical simulations to show that the $\approx 10^7$ K emitting gas is likely to be produced by stellar feedback in and around the Galactic disk. We further show that the emitting gas at both super-virial and virial temperatures in the extraplanar regions is metal enriched and is not in hydrostatic equilibrium with the halo but is continuously evolving. 
\end{abstract}



\section{introduction}
\label{sec:introduction}
The gas surrounding the stellar disk and the Interstellar medium (ISM) of a galaxy is referred to as the `Circumgalactic medium' (CGM) (for a review see \citet{Tumlinson2017}, \citet{Mathur2022}, \citet{Faucher2023}). Observations of the CGM of the Milky Way (MW) and other galaxies, as well as numerical simulations, suggest the presence of multiphase gas in the CGM, with a large range in density and temperature. The gas temperature varies from the cold phase ($\sim 10^4$ K) to the virial phase ($\sim 10^6$ K). The `virial gas' ($\sim 10^6$ K or $0.2$ keV), so called because its temperature is comparable to the MW virial temperature, is the volume-filling and the most massive component of the CGM, and is detected in X-rays in both absorption and emission \citet{Gupta2012}. Gas in the cold  ($10^4$ K) and warm ($10^5$ K) phases is detected in the optical/UV band in absorption using bright background sources (typically a quasar). 

Recent observations have, however, indicated the presence of $\sim 10^7$ K ($0.8$ keV) gas in addition to the gas at lower temperatures. Following \citet{Das2021}, we refer to this $\sim 10^7$ K gas as the `super-virial' gas (instead of `coronal' or `hot' gas, to avoid confusion with the $\sim 10^6$ K gas). We will make further distinctions regarding this phase, as follows. We will refer to the `super-virial' phase that has been detected using SiXIV and NeX absorption lines in the X-ray spectra of background quasars along a few lines of sight (\citet{Das2019a}, \citet{Das2021}, \citet{McClain2023}, \citet{Lara-DI2023}), \citet{Lara-DI2024}, as `SV absorption'. And by `SV emission' we will refer to the hot gas that has been detected in the emission studies of the MW CGM in many fields across the sky, in addition to the virial phase. 
In this paper, we will focus on the `SV emission' phase, and the case of SV absorption will be dealt with in a separate paper.

 Several X-ray observatories (eg., {\it Chandra} (\citet{Weisskopf2000}), {\it XMM-Newton} (\citet{Jansen2001}), {\it Suzaku} (\citet{Mitsuda2007}), {\it HaloSat} (\citet{Kaaret2019}), {\it eROSITA} (\citet{Predehl2021})) have been put into service for these emission observations. \citet{Das2019b} detected the super-virial phase in the MW CGM using XMM-Newton observation towards the Blazar IES 1553+113 ($\it{l} = 21^{\circ}.91, \it{b} = 43^{\circ}.96$) in emission. They fitted the total emission from the CGM with a two-temperature thermal plasma model. Assuming solar metallicity, their best-fitted values of two temperatures are $0.15-0.23$ and $0.41-0.72$ keV.
\citet{Gupta2021} used {\it Suzaku} and {\it Chandra} observations of the MW halo to characterize the emission along four sightlines. Again, for solar metallicity, they found the two temperatures to be $0.176\pm 0.008$ keV and $0.65\hbox{--}0.90$ keV. 
\citet{Bhattacharyya2023} used {\it XMM-Newton} observation towards the Blazar Mrk 421 (($\it{l} = 179^{\circ}.83, \it{b} = 65^{\circ}.03$))  and 5 other sightlines near the Blazar field to study the emission from MW CGM. Using solar metallicity to analyze their observation, they found the temperature of two phases to be $0.14\hbox{--}0.192$ keV and $0.66\hbox{--}1.14$ keV. 
\citet{Bluem2022} used {\it HaloSat} all-sky survey to study the emission from the MW halo. Assuming 0.3 solar metallicity, they obtained the average temperature of $0.179^{+0.005}_{-0.007}$ and $0.184\pm 0.009$ keV for northern and southern virial gas respectively. For the super-virial gas, the average temperature
is $0.69^{+0.05}_{-0.06}$ and $0.75\pm 0.08$ keV respectively.
\citet{Ponti2023} observed the MW halo in the eFEDDS field using {\it eROSITA}. 
Their best-fit metallicity was 0.068 Z$_\odot$ for the virial phase,  much less than the value of 0.3 Z$_\odot$ that is often used in CGM studies. However, they used 0.7 solar metallicity for the super-virial gas.
\citet{Sugiyama2023} studied emission from the MW Halo with {\it Suzaku} using 130 observations. They used solar metallicity in the super-virial phase for their analysis. 
Lastly, \citet{Gupta2023} analyzed the X-ray emitting gas in the CGM using $230$ \textit{Suzaku}  archival observations. Assuming solar metallicity, they found the temperature of virial and super-virial gas to be $\sim 0.2$ and $\sim 0.4-1.1$ keV respectively.
All these observations show that the super-virial hot phase of MW CGM, as observed in emission, is ubiquitous across the sky. 
While a 2-temperature model provided best fit to all these data, there is clearly some scatter in the temperature of both virial and super-virial phases, suggesting a multiphase nature of CGM.

\begin{figure}
    \centering
    \includegraphics[width=\columnwidth]{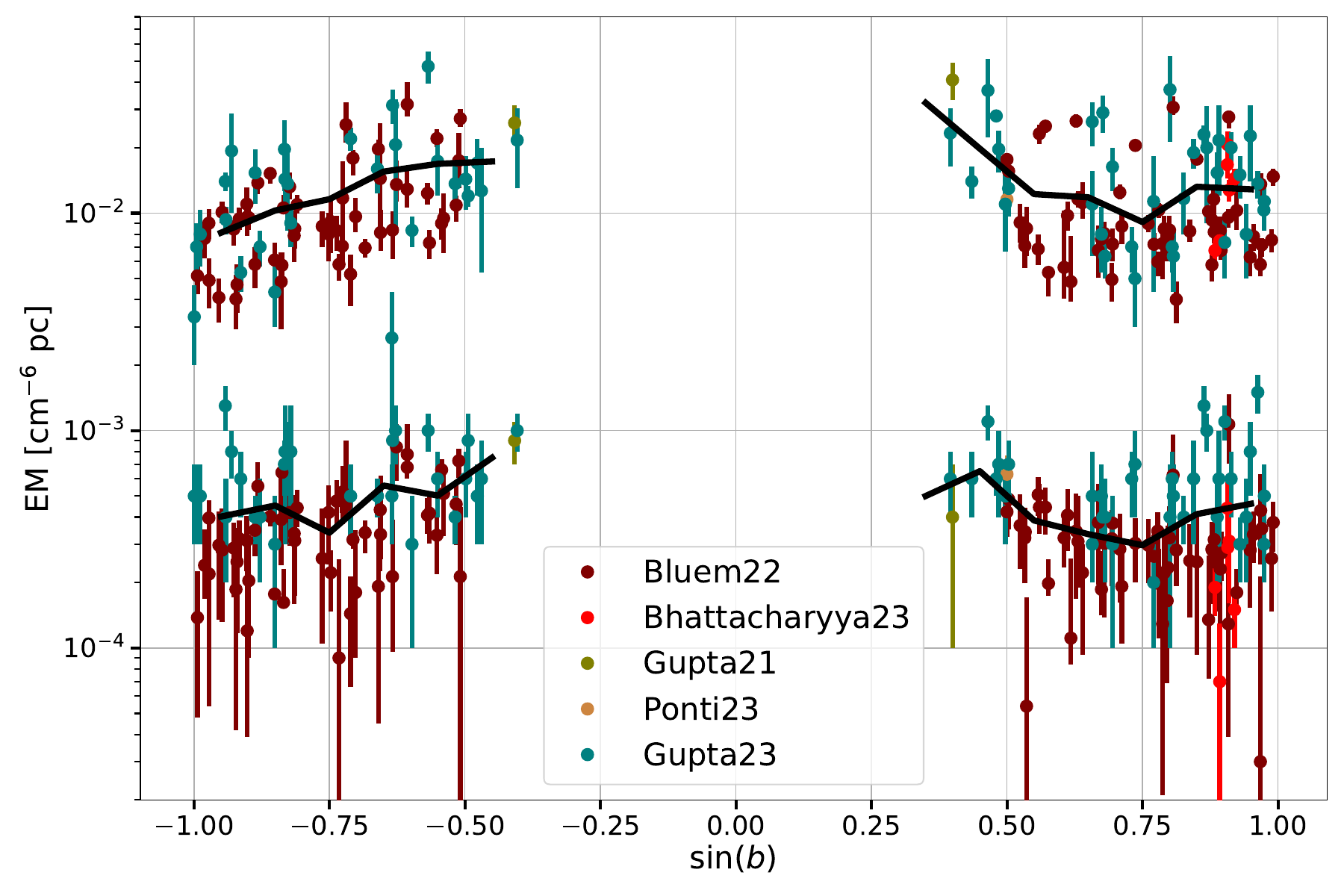}
    \caption{The Emission Measure (EM) data from various authors (\citet{Bluem2022},\citet{Bhattacharyya2023}, \citet{Gupta2021},\citet{Ponti2023} and \citet{Gupta2023}) is plotted as a function of sin$(b)$, where $b$ is the galactic latitude. The figure contains two sets of data. The top set of points, with EM $\approx 10^{-3}-10^{-1}$ cm$^{-6}$ pc, corresponds to the virial phase while lower data points with EM $\approx 10^{-5}-10^{-3}$ cm$^{-6}$ pc correspond to the super-virial phase. The solid black lines show the mean EM calculated in a bin size of $\Delta \mathrm{sin}(b)=0.1$ for both virial and super-virial datasets. Note the anti-correlation between EM and sin$(b)$ for both virial and super-virial phases as can be seen by the trend in the black solid lines.}
    \label{fig:em_sinb}
\end{figure}

Recently \citet{Fuller2023} observed and analyzed the emission from Orion-Eridanus Superbubble (OES) and found the temperature and EMs of the gas in this region to be similar to the bulk gas in the CGM elsewhere.
This result suggests that the super-virial gas is likely associated with core-collapse SNe-driven outflow from the galactic disk. 
In this paper, we model the super-virial gas ($\sim 10^7$ K) {\it and} the virial gas ($\sim 10^6$ K) with disk-like profiles, in addition to the extended CGM at the virial temperature. The unique feature of our model is the extraplanar disk-shaped region containing {\it both} super-virial and virial gas. These two gas phases need not be in hydrostatic equilibrium and are likely to be in a dynamical state. 
We explain the physical nature of the co-existence of these two phases around the disk, using an idealized numerical simulation of star-formation-driven outflows in a MW-type galaxy.
Our paper is organized as follows. Section \ref{sec:model} introduces our model, followed by a description of the simulation set-up and results in Sections \ref{sec:simulation} and \ref{sec:result}. We discuss the implications in Section \ref{sec:discussion} and finally summarise our work in Section \ref{sec:summary}.
\begin{figure}
    \centering
    \includegraphics[width=\columnwidth]{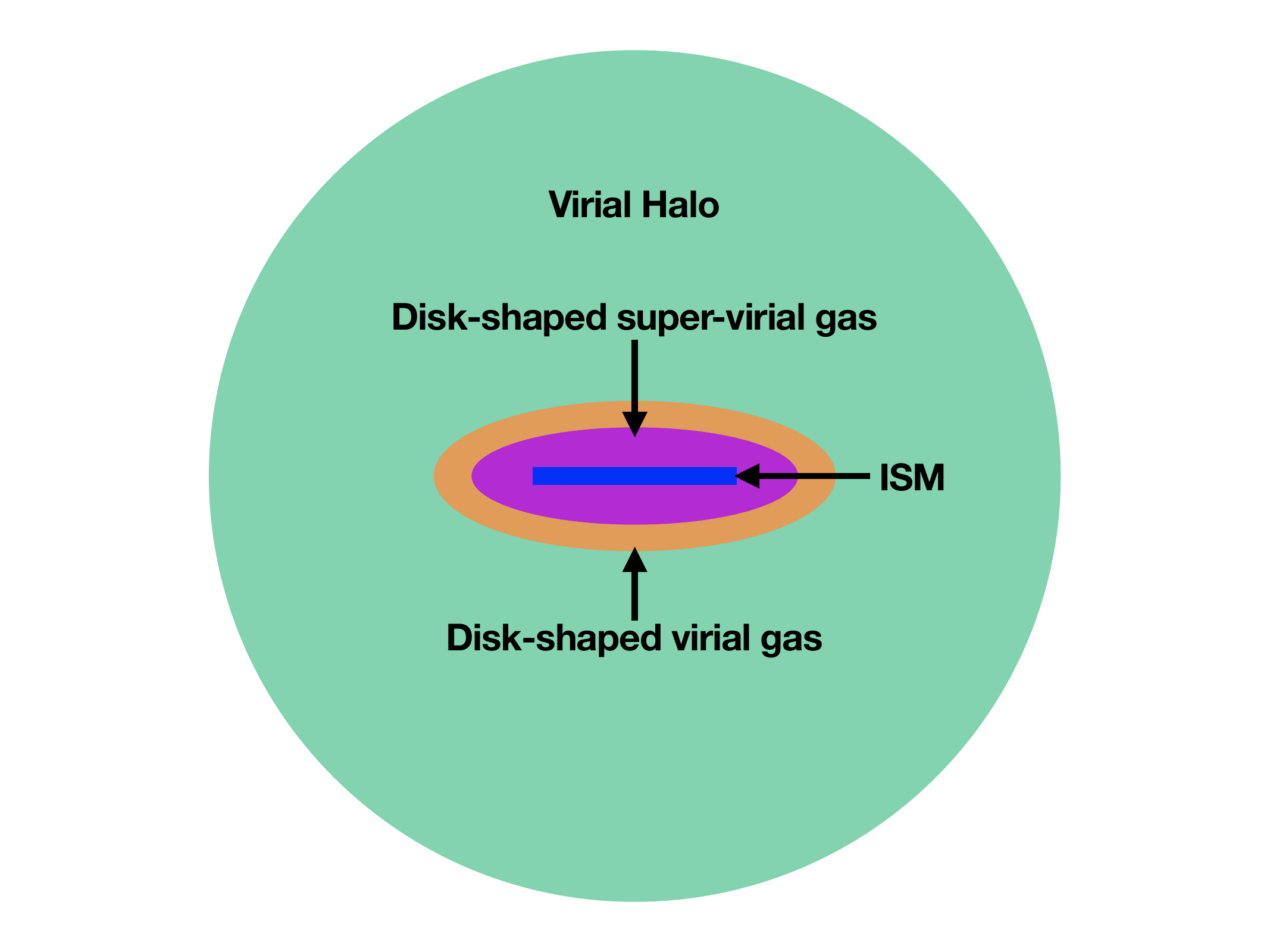}
    \caption{The cartoon representation of the structure of the gaseous components of the MW. The blue region in the center shows the Interstellar medium (ISM). The magenta and the yellow regions show the extent of the super-virial and virial temperature-emitting disks, respectively. The extended region in green shows the Galactic virialized Halo. Note that the size of different regions is not to scale; it simply represents the overall structure of the extraplanar gas around the MW.}
    \label{fig:cartoon}
\end{figure}

\begin{figure*}
    \includegraphics[width=\columnwidth]{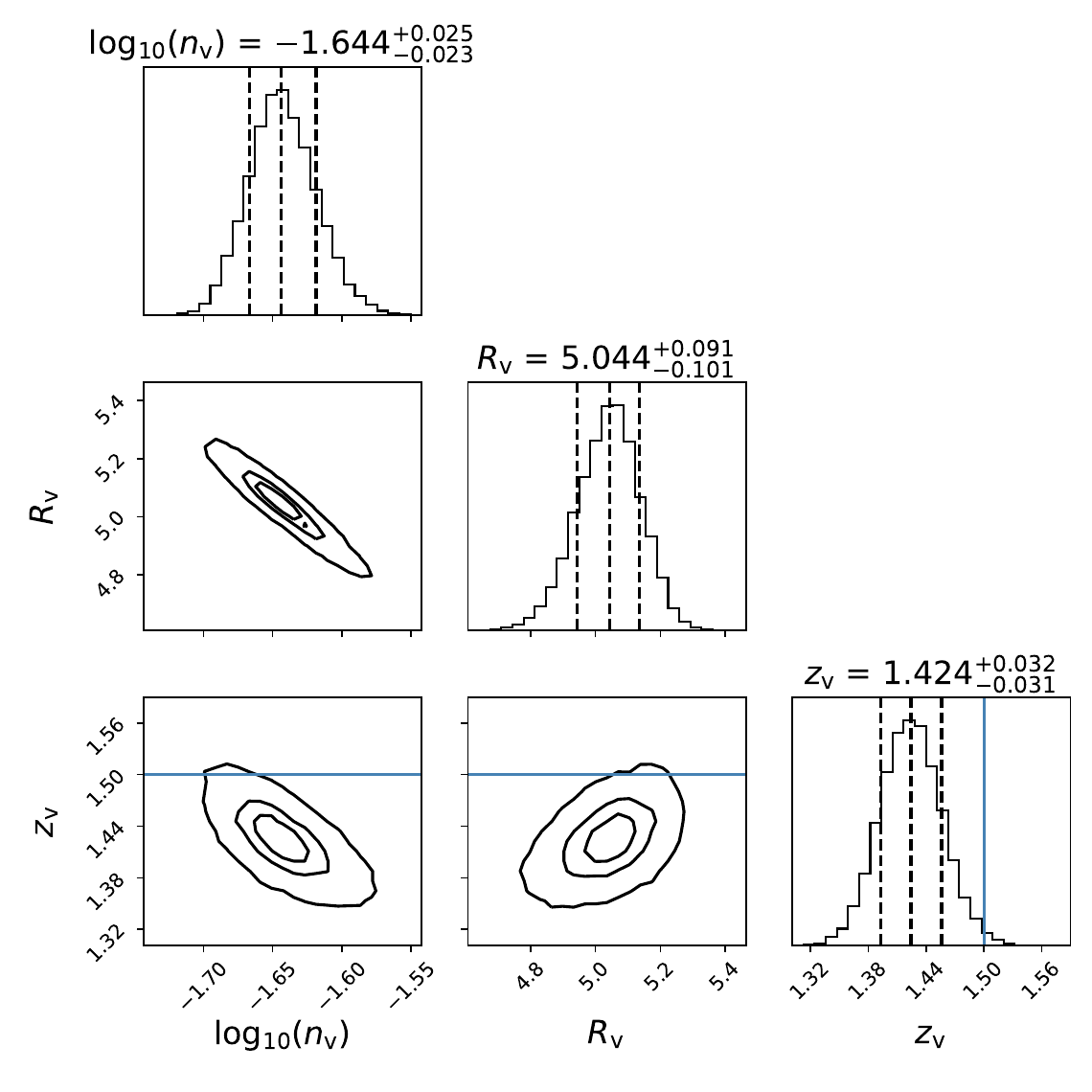}
    \includegraphics[width=\columnwidth]{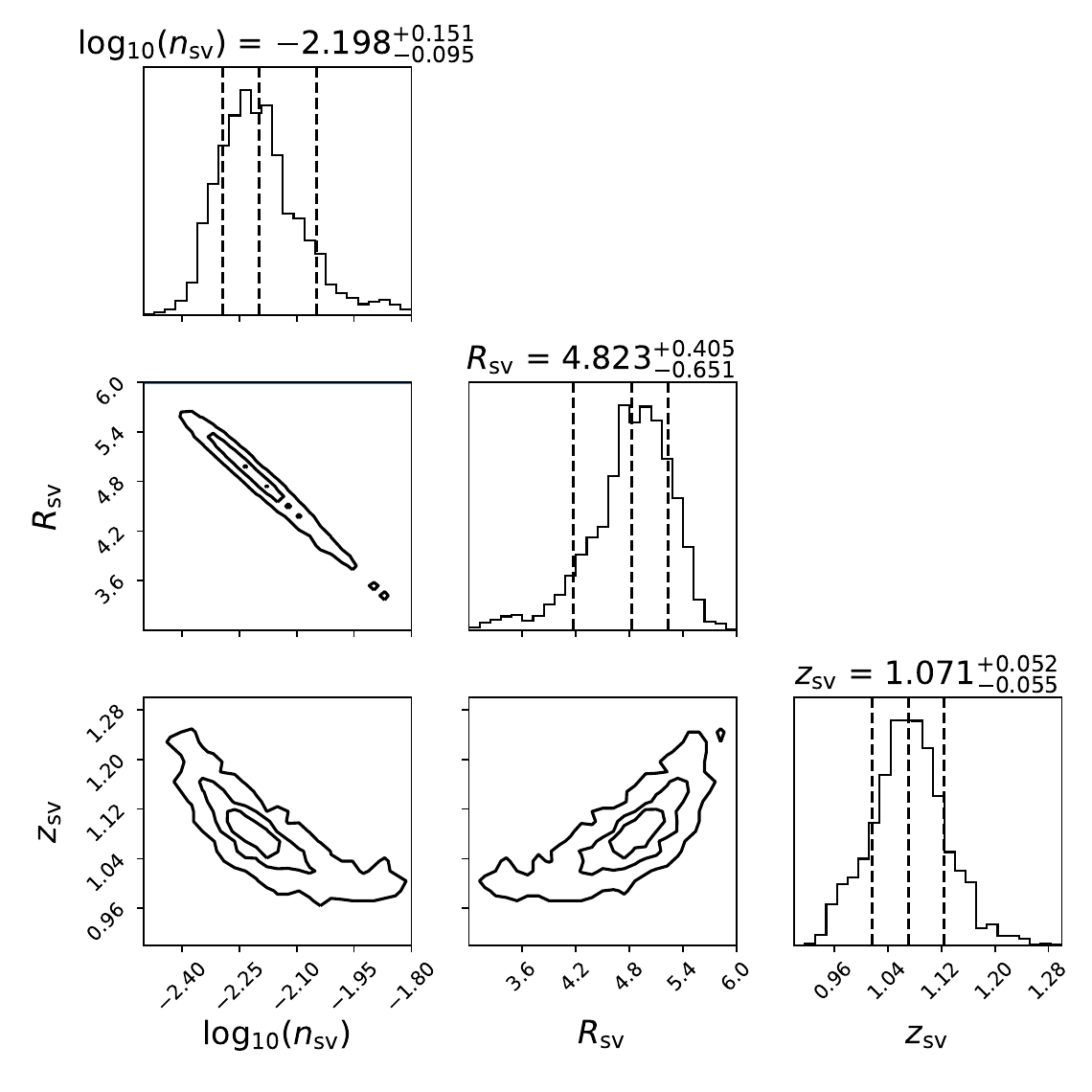}
    \caption{This figure shows the results of the MCMC analysis. The {\it left} and {\it right} panel shows the corner plot of parameters for virial and super-virial phases respectively. The three vertical dashed black lines in 1-D histogram mark the $16^{\rm th}$, $50^{\rm th}$, and $84^{\rm th}$ percentiles of the samples. The three contour levels plotted with black solid lines contain $32$, $64$, and $96$ percent of the samples respectively The blue line shows the initial guess value and is visible only in the case of $z_{\rm v}$ as the guess values for other parameters are out of the plot ranges. The values on the top of each 1-D histogram show the $50^{\rm th}$ percentile value (median) and the error shown corresponds to $16^{\rm th}$ and $84^{\rm th}$ percentile which is equal to $1\sigma$ error if the resulting 1-D histogram is Gaussian which is the case for most of the parameters. We performed the analysis using 32 walkers and 50,000 steps with Gaussian priors for all the parameters. We ensured the convergence of samples as the total number of steps are larger than 50 times the auto-correlation time. Note that the width in the Gaussian prior (see Table \ref{tab:values}) is sufficiently larger than the width in the resulting histogram of the respective parameters. This shows that the resulting values are well-constrained.}
    \label{fig:MCMC}
\end{figure*}

\section{Disk-shaped profile of the virial and super-virial gas}
\label{sec:model}
To begin with, we seek clues for the possible geometric shape of the region occupied by the `SV emission' phase. 
\citet{Kaaret2020}, \citet{Locatelli2024}, \citet{Dutta2023} and \citet{Nakashima2018} have argued that the CGM emission is best described by a disky extraplanar component. In figure \ref{fig:em_sinb} we have plotted the observed Emission Measures (EM) of the virial and the super-virial temperature gas as a function of sin$(b)$, where $b$ is the Galactic latitude for various observations. The black solid lines show the mean EM for both virial and super-virial phases calculated in bin size of $\Delta \mathrm{sin}(b)=0.1$. We see an anticorrelation of EM with sin$(b)$, as expected for a disk-shaped region, for both virial and super-virial phases as depicted by the trend in black solid lines. We have performed Kendall's tau test to quantify the anticorrelation. For the virial phase, we obtain $\tau=-0.20$ and a p-value $=10^{-4}$, where $\tau$ is the correlation coefficient and the p-value signifies the probability of the null hypothesis. $\tau/\rm p\hbox{-}value=1/0$ signifies strong correlation while $\tau/\rm p\hbox{-}value=0/1$ signifies no correlation. For the super-virial phase, we obtain $\tau=-0.18$ and p-value$=5\times 10^{-4}$. The finite negative $\tau$ values signify the anticorrelation.

\citet{Bluem2022} and \citet{Bhattacharyya2023} have shown that the EMs of the super-virial and virial phases are correlated, suggesting that they are co-spatial. Thus it appears that both virial and super-virial phase emission may arise in an extra-planar disk around the Galactic disk. A disk-like profile of the super-virial phase is also a natural outcome of the stellar activities in the disk of the MW, as we show in \S 4. However, in order to understand the virial phase absorption, an extended diffuse component is required (\citet{Gupta2012,Gupta2014}). 

With this in mind, we set up our geometric model as shown in Figure \ref{fig:cartoon}. It shows a schematic representation of our model of the extra-planar gas around the MW. The blue slab shows the Galactic ISM. The regions with magenta and yellow colors show the extent of the super-virial and virial disks respectively. The extended spherical green region shows the virial temperature CGM in the Galactic Halo. Note that the regions in the Figure are not to scale and are only schematic in nature.
In what follows, we determine the parameters of these three geometric shapes.  

First, let us consider the volume-filling diffuse gas in the Galactic halo at the virial temperature. 
Consider the halo gas in hydrostatic equilibrium with the dark matter potential of the MW,  for which we use the NFW profile (\citet{Navarro1997}). We further assume the halo gas to be at a uniform temperature of $3\times10^6$ K (isothermal sphere), for simplicity. 
Although other halo profiles are also possible (\citet{Faerman2020}, \citet{Voit2019}, \citet{Maller2004}, \citet{Miller2013}), we use the simplest model to minimize the number of free parameters. 

\begin{table*}
 \centering
   \setlength{\tabcolsep}{3pt}
   \renewcommand{\arraystretch}{1.5}
  
    \begin{tabular}{|c|c|c|c|c|c|c|c|c|}
    
    \hline
    
      \multirow{2}{4em}{Parameter} & Allowed Range & \multicolumn{2}{c|}{Initial Guess} & \multicolumn{2}{c|}{Gaussian line center} & Gaussian Half Width & \multicolumn{2}{c|}{Results}\\
      \cline{2-9}

      & V/SV & V & SV & V & SV & V/SV & V & SV \\
      \hline

     $\mathrm{log_{10}}[n$ (cm$^{-3}$)] & $[-5,-0.1]$ & $-2$ & $-3$ & $-1.5$ & $-2.5$ & $5$ & $-1.64^{+0.03}_{-0.02}$ & $-2.2^{+0.15}_{-0.10}$ \\
     \hline

    $R$ (kpc) & $[1,15]$ & $4.5$ & $6$ & $5$ & $4$ & $10$ & $5.0^{+0.1}_{-0.1}$ & $4.8^{+0.4}_{-0.7}$ \\
    \hline

    $z$ (kpc) & $[0.1,10]$ & $1.5$ & $2.0$ & $1.0$ & $1.5$ & $10$ & $1.42^{+0.03}_{-0.03}$ & $1.07^{+0.05}_{-0.06}$ \\
      
      \hline
       
    \end{tabular}
   
    \caption{Allowed Range, Initial Guess values, line center, and width in the Gaussian priors of the parameters for the MCMC analysis. The last column shows the result of the MCMC analysis for the virial (V) and the super-virial (SV) phases. The value shown are the median values ($50^{\rm th}$ percentile) with uncertainties equal to $16^{\rm th}$ and $84^{\rm th}$ percentile ($1\sigma$).}
    \label{tab:values}
\end{table*}

We obtain the following equation for the halo gas density profile
\begin{equation}
     n(r) = n_0 \ {\rm exp} \left [ -f_0\left \{ 1-\frac{{\rm ln}(1+x)}{x} \right \} \right ]\,,
    \label{eq:halo_profile}
\end{equation}

where $x=r/r_s$ ($r_s$ is the scale radius in the NFW profile), $n_0$ is the density at the centre and $ f_0 = G\mu m_p M_{\rm vir} / [k_B T r_s \{ {\rm ln}(1+C)-C/(1+C)\}]$ ($C$ being the concentration parameter). We use $M_{\rm vir} = 10^{12} \ M_{\odot}$, $C=12$, $r_{\rm vir}=260$ kpc (\citet{klypin2002}) and $T=3\times10^6$ K,  for which  $f_0 = 3$. The central density $n_0$ is obtained from the normalization condition that the total baryonic mass in the CGM ($1.2\times10^{11}\ M_{\odot}$) is $75\%$ of the total baryonic mass in MW  (\citet{Prochaska2019}) assuming cosmic baryonic fraction of $0.16$. This gives $n_0 = 8.8\times 10^{-4}$ cm$^{-3}$.

With this, we calculate the Emission Measure ($ EM = \int n_e n_H d{\it l}$) of the virial temperature halo gas. We used $\mu=0.67$, $\mu_e=1.18$ and $\mu_H=1.22$ such that $n\mu = n_e\mu_e = n_{\rm H}\mu_{\rm H} $.
We then subtract the halo EM from the total EM of the virial temperature gas so as to determine the EM from the disk component. We note that the halo contributes only $30 \%$ on average to the total EM of the virial phase. 

The disk-like model for the extra-planar profile of the virial temperature gas is given by:
\begin{equation}
    \centering
    \label{eq:02disk}
     n_{V}(R,z) = n_{V} \ {\rm exp}\left(-\frac{\lvert z \rvert}{z_{V}}\right) \times
    \begin{cases}
    1; \qquad\qquad\qquad\quad R<R_{V} \\
     {\rm exp}\left(-\left[\frac{R-R_{V}}{2}\right]\right); \ R\geq R_{V}\,.
    \end{cases}
\end{equation}

where $ n_{V}$ is the total number density at the Galactic center for the virial phase, $ z_{V}$ is the scale height, and $ R_{V}$ is the scale radius. Note that the density profile is exponential in the $z$-direction only below $ R_{V}$ while above $R_{V}$, it is exponential in both $R$ and $z$ directions where $R$ and $z$ are the cylindrical coordinates.

Similarly, we model the extra-planar super-virial gas profile with a disk-like shape, as follows:

\begin{eqnarray}
    \label{eq:08disk}
    \nonumber
     n_{SV}(R,z) = && n_{SV} \ {\rm exp}\left(-\frac{\lvert z \rvert}{z_{SV}}\right) \\
    && \times
    \begin{cases}
    1; \qquad\qquad\qquad\quad  R<R_{SV} \\
     {\rm exp}\left(-\left[\frac{R-R_{SV}}{2}\right]\right); \ R\geq R_{SV}
    \end{cases}
\end{eqnarray}

where $ n_{SV}$ is the total number density at the Galactic center for the super-virial phase, $ z_{SV}$ is the scale height, and $ R_{SV}$ is the scale radius. 

To determine the parameters of the disk profiles of both phases, we perform MCMC analysis using \textit{emcee} (\citet{Foreman-Mackey2013}) for the observed EMs reported in the literature (\citet{Das2019b}, \citet{Gupta2021}, \citet{Bluem2022}, \citet{Ponti2023}, \citet{Gupta2023} and \citet{Bhattacharyya2023}). To obtain the parameters for the extra-planar virial gas, we subtract the EM contribution from the extended halo virial gas profile (see Equation \ref{eq:halo_profile}, isothermal profile) and use the residual EM values as the observed data as discussed earlier in the section. 
We assume $0.3$ solar metallicity for the virial phase (\citet{Toft2002},\citet{Sommer-Larsen2006}, \citet{Prochaska2017}) and solar metallicity for the super-virial phase (motivated by the fact that the super-virial phase is likely the result of outflows from core-collapse SNe in the MW disk). We accordingly scale the observed EM values as $\rm EM(Z) = EM(Z_\odot) \times (Z/Z_\odot)^{-1}$. We performed MCMC analysis separately for the extra-planar virial and super-virial gas. We use Gaussian priors for all the parameters with 32 walkers and 50,000 steps. We ensured the convergence of the samples as the number of steps (50,000 in this case) is larger than $50$ times the auto-correlation time. In Table \ref{tab:values}, we show the allowed range, initial guess value, line center and width of the Gaussian priors, and the results of all the parameters in MCMC analysis for both virial and super-virial phases. Figure \ref{fig:MCMC} shows the corner plot of MCMC analysis. We find that the parameters of the extra-planar profiles for both SV and virial phases are similar, with comparable values for the scale heights and scale radii, although densities are different. The virial phase emission is significantly brighter (higher EM) compared to the SV phase, resulting in a higher density for the virial phase.

\begin{figure*}
    \centering
    \includegraphics[width=0.7\linewidth]{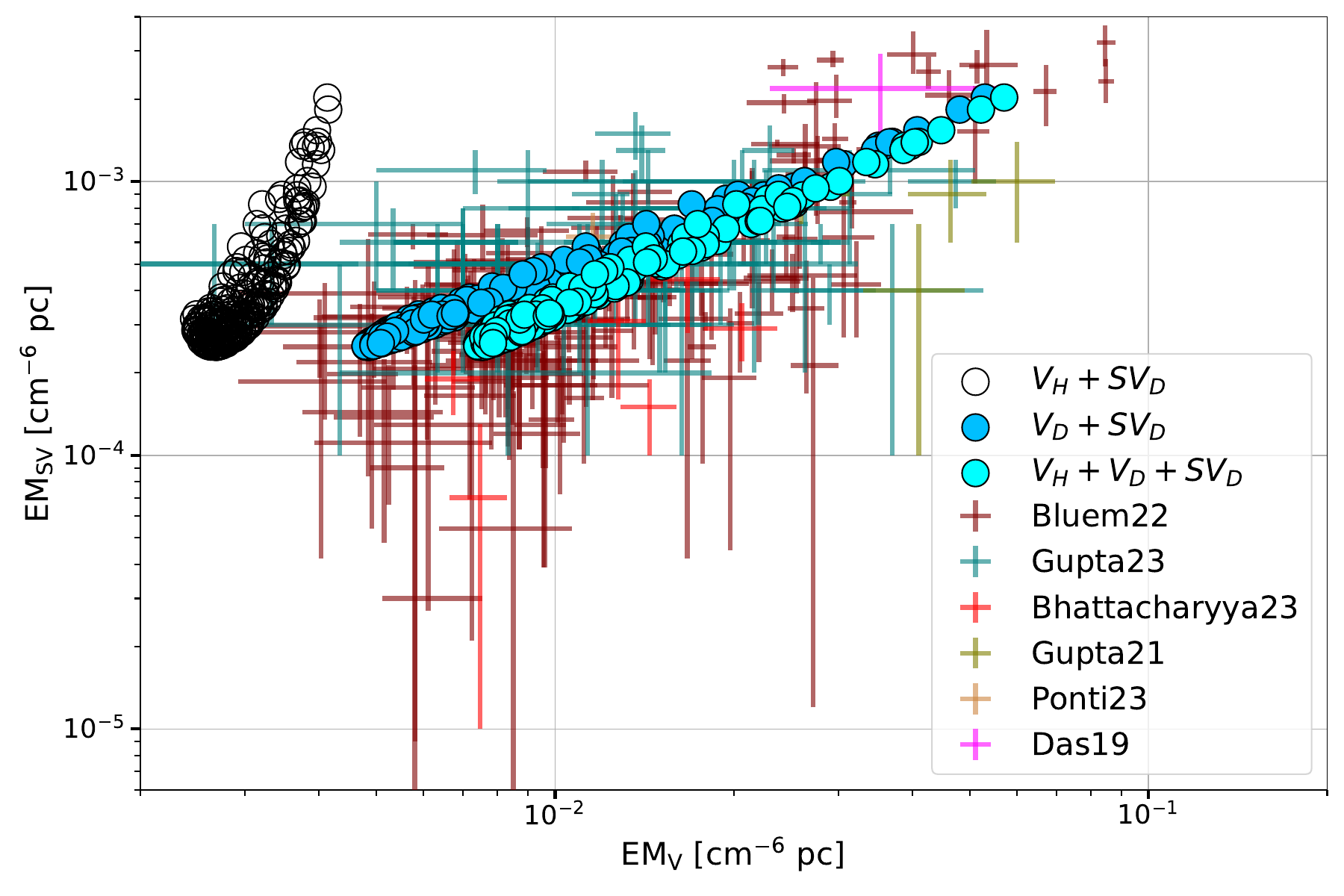}
    \caption{The correlation between the Emission Measure (EM) of the virial temperature gas and the EM of the super-virial temperature gas based on our models overplotted on observed EM. The black open circles show the EM correlation for virial halo profile and super-virial gas in the disk. The blue circles show the EM correlation with the virial and super-virial gas around the disk. The circles in cyan correspond to the virial gas in the halo + the virial gas in the disk and the super-virial gas in the disk. The EM values from our model are plotted for the observed lines of sight. The observational data points from various authors are shown in other colors with error bars. All the observed EM values are rescaled to the metallicity of $0.3$ solar for virial gas and solar for super-virial gas. The narrow range in the black points compared to cyan points along the X-axis shows the importance of having a `disk-shaped' virial phase profile.}
    \label{fig:emcor}
\end{figure*}

Using the results of the MCMC analysis for the disk-shaped virial and super-virial gas and the extended volume-filling halo virial gas using Equation \ref{eq:halo_profile}, we calculated the EMs for both virial and super-virial phases along the observed lines of sight. Figure \ref{fig:emcor} shows the EM correlation of the two phases. The black open circles show the correlation with the halo virial profile along with extra-planar super-virial gas. The blue circles correspond to the extra-planar virial and super-virial gas (and no halo gas). The circles in cyan show the correlation in the presence of a halo virial, extra-planar virial, and extra-planar super-virial gas. The observed EM from various authors are plotted with different colors; \citet{Bluem2022} in maroon, \citet{Gupta2023} in green, \citet{Bhattacharyya2023} in red, \citet{Gupta2021} in olive, \citet{Ponti2023} in peru and \citet{Das2019b} in magenta.

It is clear from Figure \ref{fig:emcor} that the model without extra-planar virial gas (black open circles) cannot explain the observed data; a disk-like extraplanar region with virial temperature gas is required. 
The observed correlation is well explained by the extra-planar super-virial and virial gas model with an extended virial halo profile (cyan-colored circles). 
The total mass in the disk-like extra-planar region  (Equation \ref{eq:08disk}, \ref{eq:02disk}) is $M=10^8 \times n_0 z_0 (8+4R_0+R_0^2) \ \rm M_{\odot}$, where $n_0$ is the central density in $\rm cm^{-3}$, $z_0$ and $R_0$ are the scale height and scale radius in kpc respectively. We calculated the mass in the extra-planar virial and the super-virial phase using the best-fit parameters from Table \ref{tab:values}. We obtain $\rm 1.7\times10^8 \ M_\odot$ and $\rm 3.5\times10^7 \ M_\odot$ for the two phases respectively. Note that this is insignificant compared to the mass of the virial gas in the halo, and, therefore, does not affect the hydrostatic equilibrium that was assumed to obtain Equation \ref{eq:halo_profile}. The baryonic mass of the CGM is dominated by the extended, diffuse gas filling the Galactic halo. 

\section{Simulation setup}
\label{sec:simulation}

In \S2, we determined the location of the virial and super-virial phases in emission, measured the parameters of the extra-planar disks, measured the density profiles, calculated the EMs, and determined masses in these phases. This answers the question ``where is the hot gas?''. However, this does not tell us why. A likely answer is the feedback from the stellar disk. In order to understand the origin of the SV-phase gas, we have built a model of the outflows from the OB associations in the MW disk. We perform a hydrodynamical simulation of outflows from the MW disk and show that the extra-planar super-virial gas may indeed arise from the feedback. 

\begin{figure*}
   \centering
    \includegraphics[width=\textwidth]{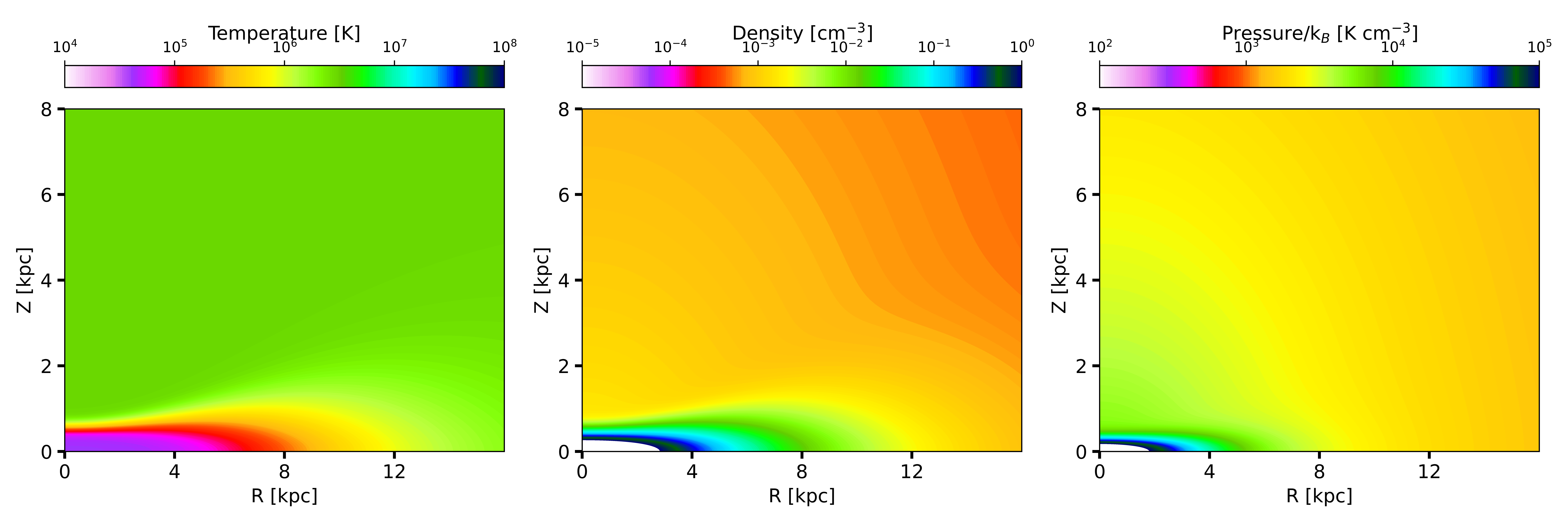}
    \caption{The initial temperature, density, and pressure maps. There are temperature and density gradients in both $R$ and $z$ directions, as a result of the hydrostatic equilibrium in which the gravitational force is balanced by gas pressure combined with the rotation of the gas. }
    \label{fig:initial} 
\end{figure*}

We performed our simulation in 2-D Cylindrical coordinates ($R$ and $z$), using the publicly available hydrodynamical code PLUTO (\citet{Mignone2007}). Our initial Galactic setup is similar to \citet{Sarkar2015} with a rotating ISM and non-rotating isothermal CGM, in the gravitational field of a Navarro-Frenk-White halo (\citet{Navarro1997}) and Miyamoto-Nagai disk (\citet{Miyamoto1975}). The temperature, density and pressure initial conditions are as shown in Figure \ref{fig:initial}. The cold ISM is at $T=4\times10^4$ K and the CGM is at $T=3\times 10^6$ K. The temperature gradient, as shown in Figure \ref{fig:initial} allows for a stable thermal configuration, for the gravitational forces to be supported by gas pressure and rotation, as discussed in \citet{Sarkar2015}. We use the HLL Riemann solver and resolve the time using the Runge-Kutta 2nd order scheme.

\textit{Grid}: We define our simulation box from $10$ pc to $30$ kpc along the $R$ direction. We take a uniform grid from 10 pc to 10 kpc (450 grid points) and a logarithmic grid from $10$ kpc to $30$ kpc (62 grid points). In the $z$ direction, we define our computational box from $1$ pc to $30$ kpc with a uniform grid from $1$ pc to $10$ pc ($20$ grid points) and then logarithmic grids from $10$ pc to $30$ kpc ($236$ grid points). 

\textit{Boundary condition}: We use reflective boundary conditions for inner boundaries in both $R$ and $z$ directions. For the outer boundary in both $R$ and $z$, pressure and density values are set to initial equilibrium values whereas velocity values are copied from the nearest active zones. We set `axisymmetric' boundary condition in the $\phi$ direction.

\textit{Cooling}: We do not allow the Galactic ISM ($R<15$ kpc and $z<2$ kpc) to cool. However, we allow the injected and CGM gas to cool, using a radiative cooling function. The metallicity for the injected material and CGM gas is set at  $1\,Z_{\odot}$ and $0.3\,Z_{\odot}$ respectively. We set the cooling to be {\it TABULATED} and use the radiative transfer code {\it CLOUDY} (\cite{Ferland2017}) to obtain the cooling rates.

\begin{figure*}
    \centering
    \includegraphics[width=\textwidth]{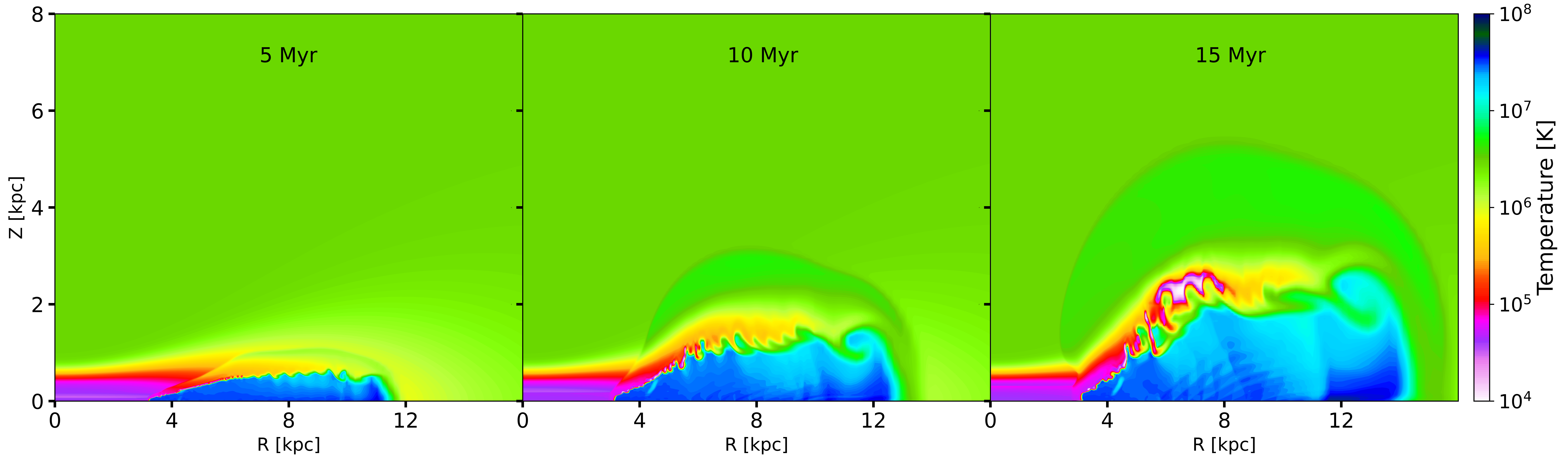}
    \caption{Temperature maps at three epochs, at $5,\, 10$ and $15$ Myr, from left to right. As the energy is injected in the disk, the hot SV-temperature gas is pushed up, generating an extraplanar region (shown in blue). Above this region, a virial temperature extraplanar region is formed (shown in yellow-green). Shock heating of the halo gas leads to another hot region at higher $z$ (shown in light green).} 
    \label{fig:time_evol}
\end{figure*}

Various studies have shown that star-forming regions of the MW are not concentrated in the Galactic center but are distributed in rings in the Galactic plane with a peak of star formation around $4\hbox{--}5$ kpc (e.g. \citet{Elia2022}). Motivated by these observations, we injected mass and energy into the Galactic disk rather than solely in the Galactic center. The star formation rate density has been measured in concentric rings of widths $0.5$ kpc by \citet{Elia2022} (see their Figure 6). For our calculation, we use these measurements in rings up to $10$ kpc. We computed the star formation rate (SFR) in each bin using the observed star formation rate density and the relevant area of the ring. The estimated SFR in the rings, starting from the center, is $0.035$, $0.018$, $0.011$, $0.016$, $0.015$, $0.038$, $0.048$, $0.093$, $0.174$, $0.201$, $0.170$, $0.158$, $0.184$, $0.125$, $0.079$, $0.074$, $0.076$, $0.068$, $0.045$ and $0.055$ $\rm M_{\odot} \, yr^{-1}$ respectively.

\textit{Energy and mass injection}: We inject mass and energy in $20$ cylindrical rings (of height $10$ pc and width $0.5$ kpc) up to $10$ kpc centered around $0.25,0.75,...9.25$ and $9.75$ kpc. The amount of mass and energy injected in each bin depends on the SFR of the bin (as mentioned above). Note that this is similar to \citet{Vijayan2018} except that we use the observed SFR in each bin from \citet{Elia2022} whereas \citet{Vijayan2018} used gas density to infer the SFR using Kennicutt-Schmidt relation (\citet{Kennicutt1998}). Assuming that each SN releases $10^{51}$ erg of energy, the relation between mechanical luminosity and SFR is,
\begin{equation}
     \mathcal{L} = 10^{51} \, \mathrm{erg} \times n_{\rm SN} \times \epsilon \times \rm SFR ( M_{\odot} \, yr^{-1}) 
\end{equation}
where $n_{\rm SN}$ is the number of SNe per unit mass of the stars formed, and $\epsilon$ is the heating efficiency of the gas. Assuming Salpeter IMF (\citet{Salpeter1955}) between lower and upper limit on stellar mass of $0.1$ and $100$ $M_{\odot}$ respectively, we get $n_{\rm SN} = 7.6 \times 10^{-3}$. Thus the relation becomes,
\begin{equation}
    \mathcal{L} = 2.4\times10^{41} \rm erg \, s^{-1} \times \epsilon \times SFR (M_{\odot} \, yr^{-1})
\end{equation}
We assume $\epsilon=0.3$, which takes into account the radiative loss of energy \citep{Vasiliev2015, Yadav2017}. This quantifies the amount of energy being injected in each bin. We inject total energy in the thermal form.

The corresponding mass injection rate is quantified by the `returned fraction' of stellar mass that is returned to the ISM during the stellar evolutionary process. We use a value $0.1$ for this returned fraction \citep{Tinsley1980} and inject mass in each bin with the rate $\dot{M}_{\rm inj}=0.1 \times \rm SFR$. We continuously inject mass and energy for $30$ Myr.

\begin{figure*}
    \centering
    \includegraphics[width=0.8\textwidth]{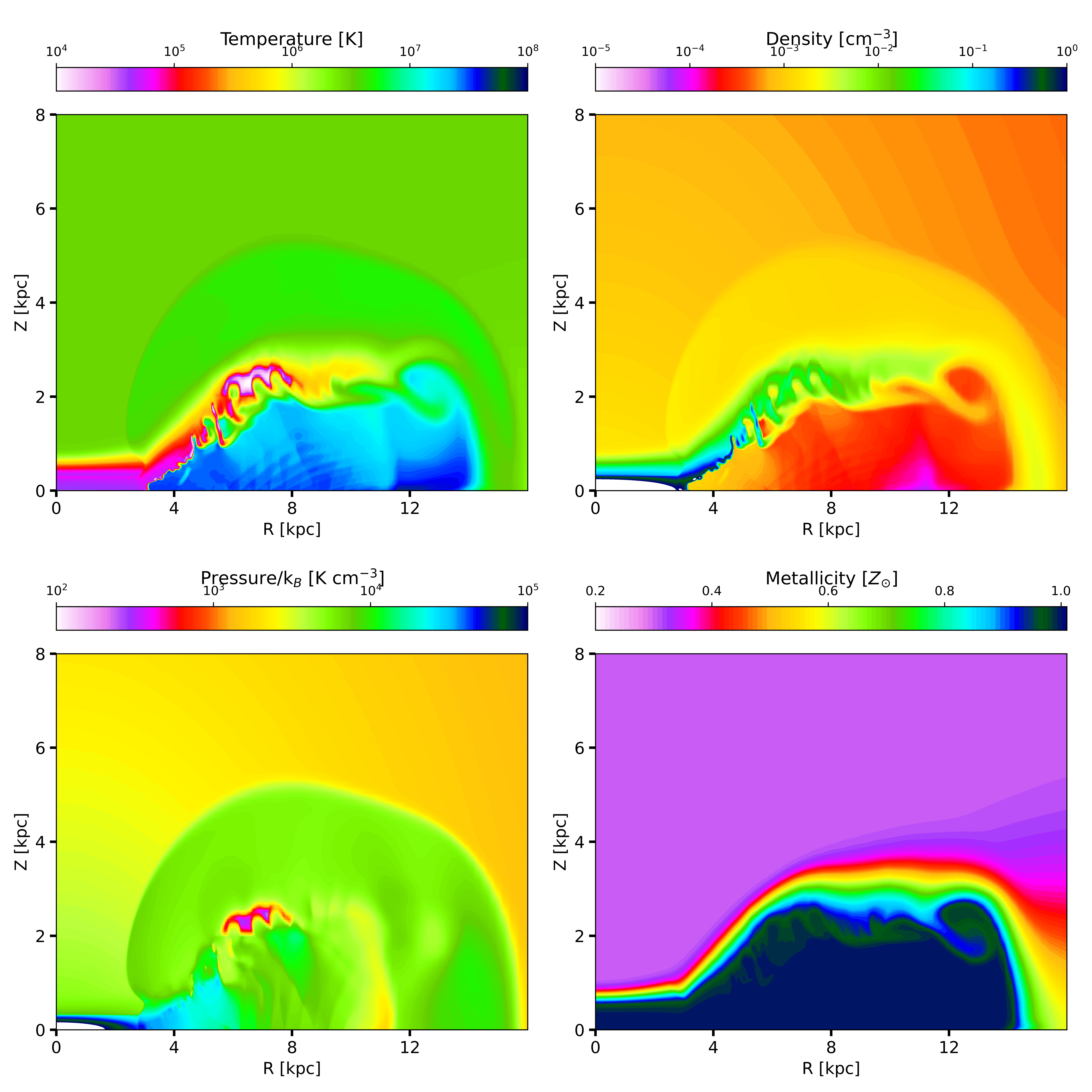}
    \caption{This figure shows the simulation results at 15 Myr. The \textit{top left} and \textit{right} panels show the temperature and particle number density maps respectively. The \textit{bottom left} and \textit{right} panels show the pressure and metallicity maps respectively.}
    \label{fig:simulation}
\end{figure*}

\section{Simulation Results}
\label{sec:result}
We post-process the simulation data to characterize the extra-planar super-virial and virial gas.
In Figure \ref{fig:time_evol}, we show the temperature evolution at three epochs ($5, 10$, and $15$ Myr). 
For the Salpeter Initial Mass Function (IMF) (\citet{Salpeter1955}), the average mass of supernova progenitors is $\approx 18$ M$_\odot$ (using a threshold of $8$ M$_\odot$ for SN progenitors, and stellar masses being distributed from $0.1$ to $100 \ \rm M_{\odot}$). The main sequence lifetime of a $18 \rm M_{\odot}$ progenitor star is $\sim 15 \, \rm Myr$. In other words, the typical time scale for the effects of star-formation-driven perturbations to manifest themselves is $\approx 15$ Myr. This estimate has motivated our choice of epochs to be showcased here. 

The blue region in Figure \ref{fig:time_evol} shows the hot injected gas with super-virial temperature ($T\geq 5\times 10^6$ K); as time evolves, the SV gas phase extends to higher $z$ values. The extra-planar region at the SV temperature is thus formed, with a height of about $1$ kpc at $10$ Myr and about $2$ kpc at $15$ Myr.

As the injected gas pushes up from the ISM, it heats the gas initially at $T=10^5$ K (red region in the left panel of Figure \ref{fig:initial}), eventually reaching $T\approx 10^6$ K, in $10$ Myr (yellow--green region in Figure \ref{fig:time_evol}. We identify this extraplanar region with the disk emitting at the virial temperature. At even larger heights, we see a `light green region' at $T=(4-5)\times 10^6$ K. This region is discussed further in \S 4.1. 

Figure \ref{fig:simulation} shows the details of our results at $15$ Myr with temperature, particle number density, pressure, and metallicity maps.  
As in Figure \ref{fig:time_evol}, we see that the SV hot phase is extended out to about $2$ kpc (light blue region in the top left panel). The density in this region is $\approx 10^{-3}$ cm$^{-3}$ (as seen in the top right panel). The region above this SV-phase, identified with extraplanar region with virial temperature (yellow-green region in the top left panel), has a higher density of $\approx 10^{-2}$ cm$^{-3}$ (green region in the top right panel). 

The bottom left panel shows the pressure map. We clearly see the high pressure regions beyond $R\approx 4$ kpc, showing outflow from the disk. The final pressure is significantly higher than the initial pressure, showing that the virial and super-virial gas in the disk is out of hydrostatic equilibrium and dynamically evolving. Note that we do not see strong outflows from the inner regions ( $0\hbox{--}3$ kpc) of the disk as compared to outer regions. The reason is that the SFR has a Gaussian-like profile along the disk with peak SFR occurring around $\approx 4\hbox{--}5$ kpc. The other reason is that the ambient density in the inner regions is higher than that in the outer regions (see middle panel of Figure \ref{fig:initial}). The high density in the inner regions provides high resistance to the outflowing gas whereas the outflow can penetrate the low density environment in outer regions for approximately the same SFR.

The bottom right panel shows the metallicity map. The enriched and high metallicity gas in the outflow mixes with the low metallicity ambient gas thus contaminating the chemical composition of the injected gas. We have used tracers for the CGM gas and the injected gas to trace the corresponding metallicity. Tracers follow the advection equation for the respective gases. Thus the tracer tracks the gas and accounts for the amount of mixing of the gases with different tracer values. The tracer value in the mixed gas is the mass-weighted value of the individual gas tracers, thus it can be used as a proxy for metallicity. Recall that the CGM metallicity is assumed to be $0.3 \, Z_\odot$  while the injected material has solar metallicity. The map 
shows that the outflow lifts the high metallicity gas out to the extraplanar region with a height of about $2$ kpc; this is the region where we have the SV hot gas. Thus we see that the hot gas is metal enriched. Beyond the extraplanar region, we see the mixing of the high metallicity gas from the outflow and the low metallicity CGM gas, with the average metallicity decreasing with $z$. Beyond $z\approx 4$ kpc, we see the low-metallicity of the CGM, without any mixing from the outflow. 

\subsection{Hot gas in the extra-planar region}
The temperature map presented in Figure \ref{fig:simulation} shows the gas in the extra-planar region to have a range of temperatures. It is also instructive to compare the metallicity map (bottom-right panel of the same Figure) with the top-left panel in order to understand the origin of gas with various temperatures within this extra-planar region. Recall that the initial gas profile has warm ($\sim 10^5$ K) gas hovering above the disk, joining smoothly into the virial gas in the halo (Figure \ref{fig:initial}). This parcel of gas is seen in red color in the metallicity map. As the collective gaseous disturbances triggered by SNe rise above the disk, this warm gas gets heated to temperatures comparable to that of the virial gas in the halo, as shown by the light-green colored gas in the temperature map. This likely corresponds to the extra-planar virial temperature gas discussed in Section \ref{sec:model}. A similar temperature structure of the outflowing gas from a disk-wide star formation process was also seen in \citet{Vijayan2018}.

The temperature map (top-left panel of Figure \ref{fig:simulation}) also shows an additional hot region at $z\approx 4\hbox{--}6$ kpc with $T=(4-5)\times 10^6$ K (shown with light green colour). We see in the bottom right panel Figure \ref{fig:simulation} that this region has the original metallicity of the CGM, with no mixing from the outflow. Thus the hot gas in this region is not the hot SN-driven outflowing gas. We find that this phase has been created by mild shock heating, as the outflowing gas hits the virial gas in the halo, with speed $\approx 300\hbox{--}400$ km s$^{-1}$ (note the advancement of the outflow in $z$-direction between the snapshots shown in Figure \ref{fig:time_evol}). 
Given the virial temperature of the halo gas, the shock has a low Mach number (of order $\approx 2$). The corresponding density jump is $\approx 2\hbox{--}3$; this is consistent with the density variations seen in the top-right panel  Figure \ref{fig:simulation}.
The corresponding temperature jump is also by a factor of $\approx 2$, as seen in the temperature map. 

\section{Discussion}
\label{sec:discussion}

For our fiducial model of the halo density profile, we have taken $M_{\rm vir}=10^{12} \, M_{\odot}$, $C=12$ (\citet{klypin2002}) and $T=3\times 10^6$ K. However, there is a scatter around these values. In order to determine how these parameters affect our results of the disk parameters (scale height, scale radius, and central density), we varied the halo parameters. 
We considered the following range in these parameters; $M_{\rm vir}=(0.9-1.2)\times 10^{12} \, M_{\odot}$, $C=10-14$ and $T=(2-3)\times 10^6$ K. As a result of these variations, the scale height changes by $+0.4/-0.3$ kpc from the fiducial value of $1.4$ kpc. The scale radius changes by $+0.1/-0.9$ kpc from fiducial value of $5.0$ kpc. The central density changes by $+0.8/-0.5 \, (\times 10^{-2}$ cm$^{-3}$) from the fiducial value of $2.3\times 10^{-2} \, \rm cm^{-3}$. As such, the variation in the halo parameters does not change the results of the disk parameters significantly. 

To test the convergence of our simulation, we performed a higher-resolution run. We doubled the resolution compared to the resolution of ($ R\times z$) $512\times 256$ in our fiducial run. We found that the main results show essentially no difference, signifying convergence; they were similar to the fiducial runs shown in Figure \ref{fig:simulation}. 

With our model, we have demonstrated the presence of extra-planar super-virial and virial temperature gas around the Galactic disk. The model also has higher density in the virial temperature disk than in the SV disk, as observed (\S 2), and the likely high metallicity of the super-virial gas. 
In the initial discovery papers, the CGM was modeled as a two-temperature plasma in both absorption and emission studies (\S 1). Over the years it was realized that the CGM actually has a range of temperature even in the X-ray band (\cite{Bhattacharyya2023, McClain2023}). The \sout{toy} model we present here naturally explains this range of temperature. 
We have shown that the CGM in the vicinity of the disk has these characteristics for $\sim 5\hbox{--}15$ Myr, which is the time-scale of manifestations of SNe-triggered outflows after the onset of star formation. It then follows that such is likely to be the case even in the case of self-sustained star formation in the disk for a much longer time. 

In this paper, we have provided a possible explanation for the super-virial phase that has been detected in emission studies. As noted in the Introduction, a few lines of sight have also detected the hot gas in absorption. The observed Oxygen column densities, are $(0.8-5.7)\times10^{17}$, $(4.9-12.9)\times10^{17}$, and $\rm (2.2-6.8)\times10^{16}\ cm^{-2}$ along IES 1553+113 (\citet{Das2019a}), Mrk 421 (\citet{Das2021}), and NGC 3783 (\citet{McClain2023}) respectively. Note that, we calculated the Oxygen column density from the observed column density of OVII and OVIII assuming Collisional Ionization Equilibrium (CIE) at the observed temperatures along the respective lines of sight. Can we explain the column density of the absorbing hot gas by our model? To answer this question, we estimated the Oxygen column density along three observed lines of sight using our disk-like model for the super-virial phase. The model predicted total column densities of $1.1\times 10^{19}$, $3.8\times 10^{18}$ and $1.2\times 10^{19}$ cm$^{-2}$ along IES 1553+113, Mrk 421 and NGC 3783 respectively.

Therefore, to match the predicted and observed column densities, we need the Oxygen mass fraction ($f_O$) of at least $0.2$, and $0.05$, along IES 1553+113 and NGC 3783 respectively.
This is much higher than the Solar Oxygen abundance of $4\times 10^{-4}$ (\citet{Asplund2009}).
Incidentally, the Oxygen mass fraction in the typical supernova ejecta is $\sim 0.1$ (\citet{Nomoto2006}), which is comparable to that required to explain the observations along two sightlines. However, the Oxygen mass fraction is likely to be lower due to the mixing of the gas from supernovae ejecta ($\alpha$-enriched) and the gas in the CGM with sub-solar metallicity.
The Oxygen column density along Mrk 421 is so high that it requires exceptional explanation and our disk model cannot explain such high column density. To conclude, it is difficult to explain the absorption observations with our model, and therefore, the origin of the `SV emission' phase is likely to be different than that of the `SV absorption' phase, and they should be treated separately. 

\section{Summary}
\label{sec:summary}
In this work, we have proposed a new model for the origin and extent of the gas emitting at the super-virial temperature ($\sim 10^7$ K) in the CGM of the Milky Way. We suggest that the origin of the super-virial gas observed in emission is connected to outflows from star-forming regions in the MW disk. A disk-like profile is thus naturally expected from the extended outflows in the disk. Moreover, we posit that this super-virial disk profile co-exists with a disk-shaped profile of virial gas. Together they can explain the Emission Measure (EM) correlation of the virial ($\sim 10^6$ K) and the super-virial phase ($\sim 10^7$ K) (see Figure \ref{fig:emcor}). 

To support our model, we have performed a hydrodynamical numerical simulation of a MW-type disk in which star-forming regions are located. Our simulation results show that the extra-planar super-virial gas has roughly a `puffed-up' disk-like feature, along with a similarly shaped region for the virial gas. 
Moreover, the composition of the super-virial gas is shown to be unmixed with the halo CGM gas and is thus likely to retain its high metallicity signature of SNe-enriched gas which are at the root of this super-virial phase. 

\section*{Acknowledgements}

We thank anonymous referee for the constructive comments. MSB thanks Alankar Dutta and Priyanka Singh for the useful discussion on MCMC analysis. MSB also thanks Manami Roy, Sourav Bhadra, Kartick Sarkar, and Siddhartha Gupta for clarifications with {\it PLUTO }.

SM is grateful for the grant provided by the National Aeronautics and Space Administration through Chandra Award Number AR0-21016X issued by the Chandra X-ray Center, which is operated by the Smithsonian Astrophysical Observatory for and on behalf of the National Aeronautics Space Administration under contract NAS8-03060. S.M. is also grateful for the NASA ADAP grant 80NSSC22K1121. SM is grateful for the hospitality during her visit at the Raman Research Institute during which this paper was completed. 
This research has made use of the VizieR catalogue access tool, CDS, Strasbourg, France (DOI : 10.26093/cds/vizier). 

\bibliography{reference}{}
\bibliographystyle{aasjournal}



\end{document}